\documentclass[12pt,letterpaper]{article}

\usepackage{sectsty}

\usepackage{amsmath,graphicx,lipsum}
\usepackage{afterpage,flafter,pdflscape,lscape,textcomp}
\usepackage[displaymath, mathlines]{lineno}
\usepackage[figuresright]{rotating}
\usepackage[utf8]{inputenc}
\DeclareUnicodeCharacter{00A0}{~}
\usepackage{bm,gensymb,parskip,xcolor,epstopdf,pdfpages}
\usepackage{pgfplots}
\usepgfplotslibrary{groupplots,units}
\pgfplotsset{table/search path={Tables/}}
\usepackage{footnote}
\makesavenoteenv{tabular}
\makesavenoteenv{table}
\pgfplotsset{compat=1.3}
\usepackage{tikz,subfig}
\usetikzlibrary{shapes, arrows, arrows.meta,positioning}
\tikzset{every picture/.style={/utils/exec={\sffamily}}}
\usepackage[margin=0.75in]{geometry}
\usepackage[square,numbers,sort&compress]{natbib}
\usepackage{tabularx,makecell,placeins,multirow,sansmath}
\usepackage[most]{tcolorbox}
\usepackage[roman]{parnotes}
\usepackage[toc,page]{appendix}
\usepackage{listings}
\usepackage{enumitem}
\usepackage{soul}
\soulregister\citep7
\soulregister\citet7
\soulregister\ref7
\usepackage[stable,bottom]{footmisc}

\usepackage[colorlinks,citecolor=red,linkcolor=red,urlcolor=blue]{hyperref}
\graphicspath{{Figures/}}
\usepackage[blocks,auth-sc-lg]{authblk}

\usepackage[labelfont=bf,font=sf]{caption}

\title{\LARGE \bfseries Capillary rise in vuggy media} 

\author{Hasan J. Khan\thanks{hasanjk@utexas.edu}}
\affil{Department of Geology,\\ University of Illinois at Urbana-Champaign, Urbana,  IL}
\author{Ayaz Mehmani}
\affil{ PoreMatters LLC, Houston, TX }
\author{Ma{\v s}a Prodanovi{\' c}}
\author{David DiCarlo}
\affil{Hildebrand Department of Petroleum Engineering, \\ The University of Texas at Austin, Austin, TX}
\author{Dayeed J. Khan}
\affil{Ameredev II LLC, Austin, TX}

\date{}

\begin{document}
\maketitle

\begin{abstract}
    
Carbonates are highly heterogeneous formations with large variations in pore size distribution and pore space topology, which results in complex multiphase flow behavior. Here we investigate the spontaneous imbibition behavior of fluids in vuggy carbonates.  Glass beads of 1.0 mm diameter, with dissolvable inclusions, are sintered to form multiple configurations of heterogeneous vuggy core with variations in matrix porosity, vug size, vug spatial location, and number of vugs. The core fabrication process is repeatable and allows the impact of vug textural properties to be investigated in a controlled manner. 

Capillary rise experiments are conducted in these proxy vuggy carbonate core and compared with the homogeneous non-vuggy core as reference. Continuous optical imaging is performed to track the position of the air-water interface in the cores. To understand the change in capillary height in the presence of a vug, a volume-of-fluid two-phase numerical simulation is performed in a parallel set of connected and disconnected tubes. Finally x-ray tomography scans are performed to identify the shape of the air-water interface in a select few cores. 

The results can be summarized as follows: disconnected vugs result in higher capillary rise compared to non-vuggy porous media. The vugs act as capillary barriers, diverting fluid flow to the adjacent connected channels, which ultimately results in a higher overall capillary rise. 
The results of this work highlight that radius of spontaneous invasion of aqueous phases, such as fracture fluid and hazardous wastes, are affected by vug porosity but not their distribution.

\end{abstract}

\section{Introduction}
Carbonate rocks are highly heterogeneous, complex rocks with a variety of pore types, pore shapes, and a wide pore size distribution \citep{folk_practical_1959,murray_origin_1960,choquette_geologic_1970,newberry_analysis_1996,ahr_confronting_2005,schon_chapter_2011,lucia_micropores_2013}. The pore size can vary over multiple length scales \citep{song_determining_2000}: the intergranular pores comprise the primary porosity in carbonate rock while dissolution, mineral replacement, and recrystallization \citep{lucia_rock-fabric/petrophysical_1995,zheng_carbonate_2000,coniglio_dolomitization_2003} can lead to secondary porosity on a localized level \citep{akbar_snapshot_2000}.

Large disconnected pore spaces, such as vugs, are formed by dissolution and recrystallization of the carbonate rock fabric. \citet{lucia_petrophysical_1983} showed that though isolated vugs significantly increase the porosity, they have negligible effect on the overall rock permeability. Fluid flow is controlled by the matrix pore size, and the presence of disconnected vugs fails to improve the capillary entry pressures \citep{coalson_subtle_1994}. Furthermore, these display a higher Archie cementation exponent, and can act as capillary barriers to flow. \citet{mehmani_quantification_2017} saw a similar behavior in fractures, where the fluid by-passes the fractures and preferentially flows through the matrix. Spontaneous imbibition study for these can improve the understanding of how far the fracturing fluid permeates during the soaking time, and can give an indication about the spatial location of the non-recovered frac fluid \citep{lan_water_2014}.

The handling of radioactive elements has been increasing with the global increase in nuclear power generation \citep{iaea_nuclear_2018}, increasing the probability of a spillage incident occurring. Furthermore, radioactive elements are found in many commercial products that end up in the junkyard at the end of their life cycle, and can potentially leach into the soil. In either case radionuclides can enter the plants through the soil, and spread further within the food chain \citep{zhu_soil_2000}. From a containment perspective it is pivotal to know the radius of contamination. For a carbonate rock in the presence of a vug, would the vug be filled and limit the radius of damage? Or would the fluid by-pass the vug and penetrate deeper inside the formation?

Micromodels, which are quasi-2D physical models with the pore structure etched on silicon wafers \citep{wan_improved_1996}, have been extensively used to study 2-phase pore-scale flow mechanisms in porous media \citep{wardlaw_effects_1980,dullien_hydraulic_1986,owete_flow_1987,fenwick_three-dimensional_1998}. We believe that the quasi-2D micromodels will not be able to capture the capillary flow in presence of a vug, and therefore we have fabricated synthetic glass bead cores with dissolvable inclusions \citep{khan_effect_2019} for these experiments.

In this paper we investigate the effect of vug(s) on core rock physics: we generate a set of synthetic glass bead cores with vug(s), experimentally determine petrophysical parameters for these core, observe the behavior of vug(s) on spontaneous imbibition, and contrast with the \citet{washburn_dynamics_1921} solution. Furthermore, we do a computational fluid dynamics (CFD) simulation of spontaneous imbibition in vug-pore models and compare with the experimental results. Finally we perform micro-CT scans on a select few cores and come up with our conclusions.

\section{Core preparation}

Proxy carbonate cores are fabricated by sintering 1.0 mm diameter glass beads with dissolvable inclusions \citep{khan_effect_2019}, salt wrapped in cheesecloth, placed in a graphite mold in a muffle furnace in the presence of air.The elevated temperatures soften the glass (Figure \ref{fig:tempandcore}) resulting in increased granular surface contact and ultimately a consolidated core. The cheesecloth is burned during the sintering process and the salt is dissolved by flushing the core with de-ionized (DI) water after it cools down. Alternatively, gypsum cement can be used as the dissolvable inclusion \citep{khan_replicating_2018}, which is cleared by flushing with HCl acid.

\begin{figure} \centering
\pgfplotsset{width=0.35\textwidth,height=6cm,{every axis/.append style={line width=1pt}}} 
\subfloat[Temperature profile with a peak temperature of 725 $\degree$C and exposure time of 15 mins results in a consolidated porous proxy core. \label{fig:tempandcore}]{%
\begin{tikzpicture}[font=\sffamily]
\begin{axis}[xlabel=Time (min),ylabel=Temperature (\degree C), xmin=0, xmax=300, ymin=0, ymax=800,legend cell align=left, legend pos=north east,x=0.25mm,minor x tick num=1,xtick distance=50]

\addplot [mark=none,color=red,dashed]
coordinates{(0,650) (300,650)}; \addlegendentry{\footnotesize Softening point}
\addplot [mark=none,color=black,solid] 
coordinates {(0,25) (35,550) (45,550) (57,725) (82,725) (250,25)}; 
\node[anchor=south west,inner sep=0pt] (core) at (40,10) {\includegraphics[height=33mm]{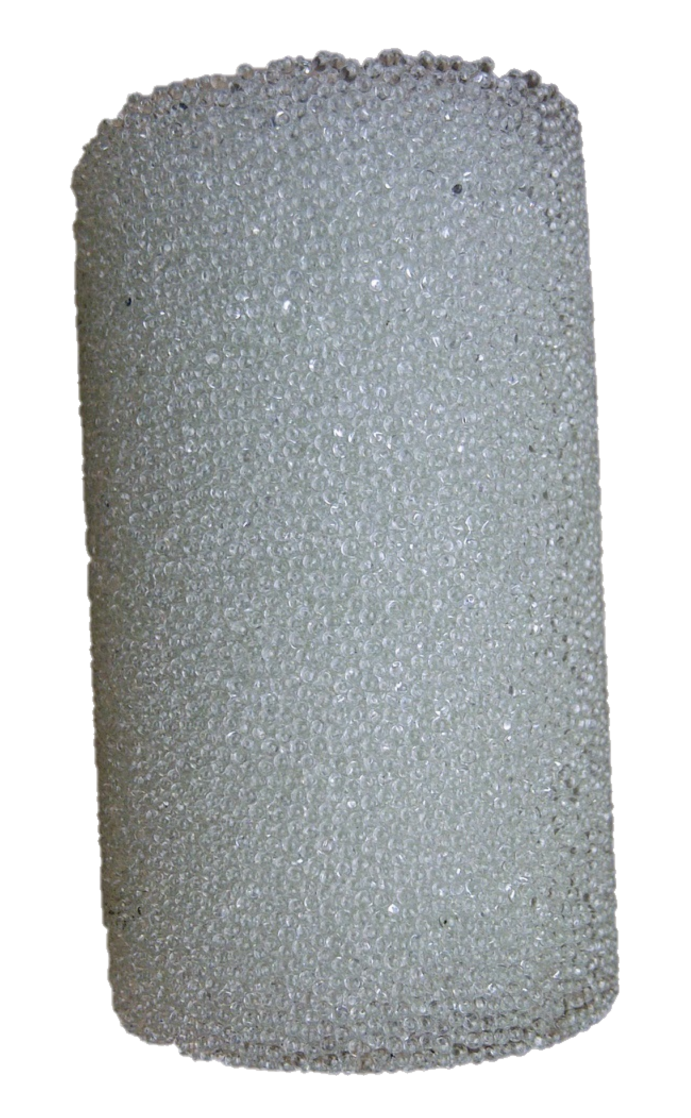}};

\end{axis} 
\end{tikzpicture} 
}
\quad
\subfloat[Pore size distribution of the fabricated core shows a matrix porosity of $\sim$41\% and average pore diameter of $\sim$195 $\mu$m. \label{fig:NMRhomo}]{%
\pgfplotsset{width=0.32\textwidth,height=6cm,{every axis/.append style={line width=1pt}}}
\begin{tikzpicture}[font=\sffamily]
\begin{axis}[axis y line*=left,xlabel={Pore diameter ($\mu$m)} ,ylabel=\textcolor{blue}{Incremental volume (ml)}, xmode=log, xmin=10, xmax=1000, ymin=0, ymax=2,minor y tick num=4,legend cell align=left, legend pos= north west,]
\addplot [blue] table [x index={0}, y index={1}]  {NMR.txt}; 
\end{axis} 
\begin{axis}[axis y line*=right, ylabel=\textcolor{red}{Porosity (\%)}, xmode=log, xmin=10, xmax=1000, ymin=0, ymax=100,xtick=\empty]
\addplot [red,dashed] table [x index={0}, y expr={\thisrow{runsum}*41.6}]  {NMR.txt}; 
\end{axis} 
\end{tikzpicture} 
}
\caption{The temperature profile results in a highly porous consolidated core with an average porosity of $\sim$41\% and an average pore diameter of $\sim$195 $\mu$m.}
\label{fig:tempandcore}
\end{figure}
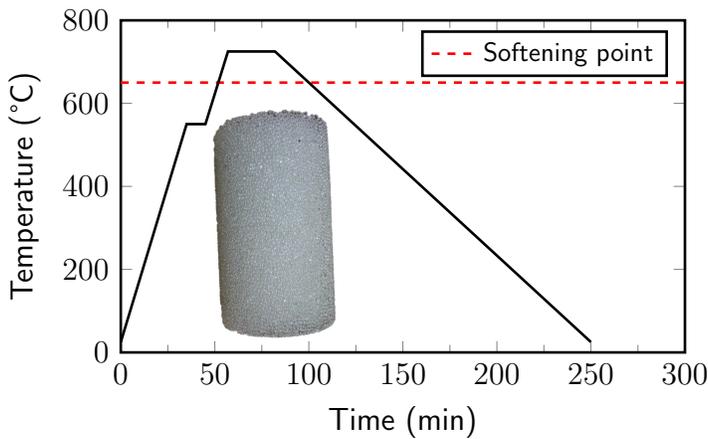
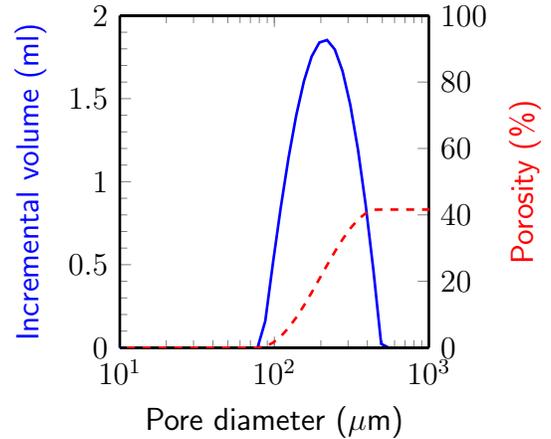

\begin{figure} \centering
\includegraphics[width=0.5\textwidth] {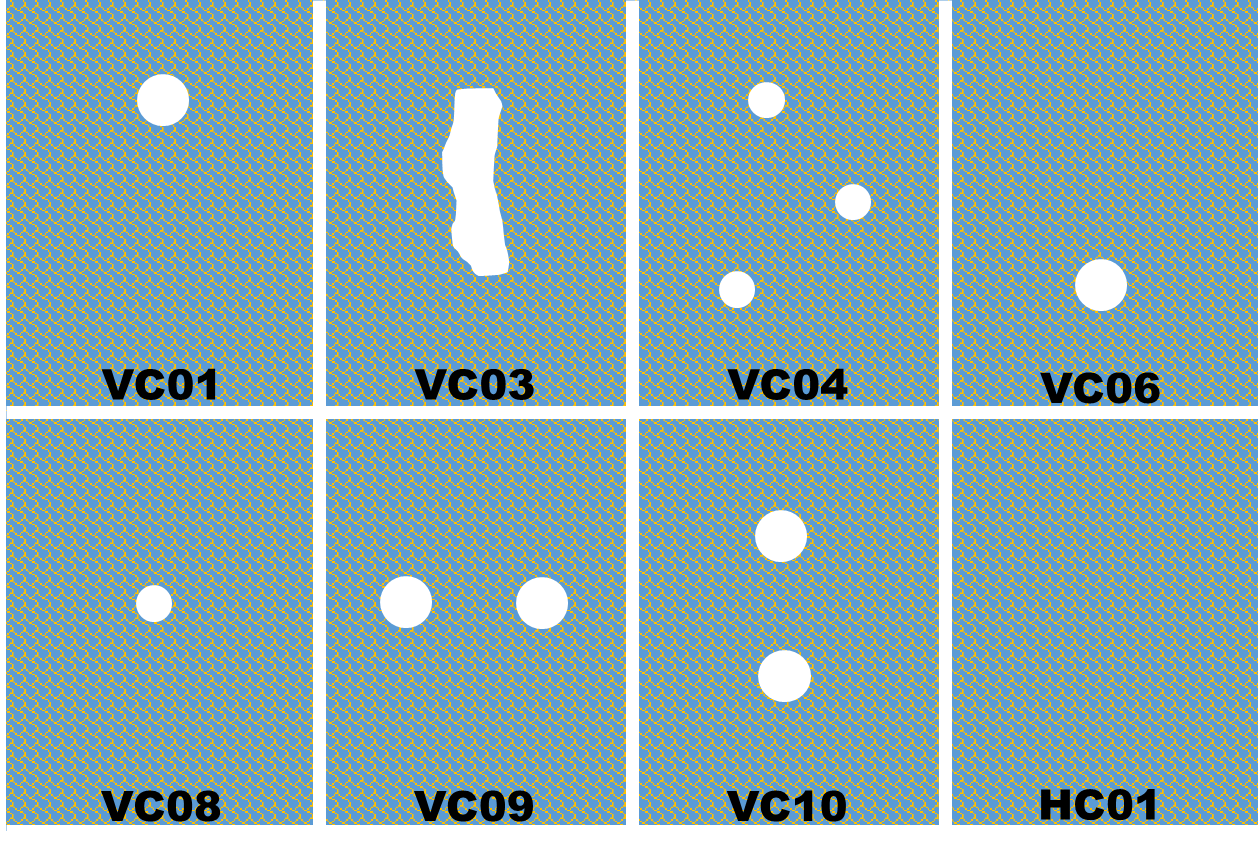}
\caption{Vug distribution (white) in the fabricated cores under study with the core name at the lower end of the core. Note that the homogeneous core is the control case and does not include any vug.}
\label{fig:vugconfig}
\end{figure}

\begin{figure} \centering
\subfloat[$\mu$CT of vug space in core VC04 (x,y,z-resolution = 29 $\mu$m).]{%
\includegraphics[width=5cm] {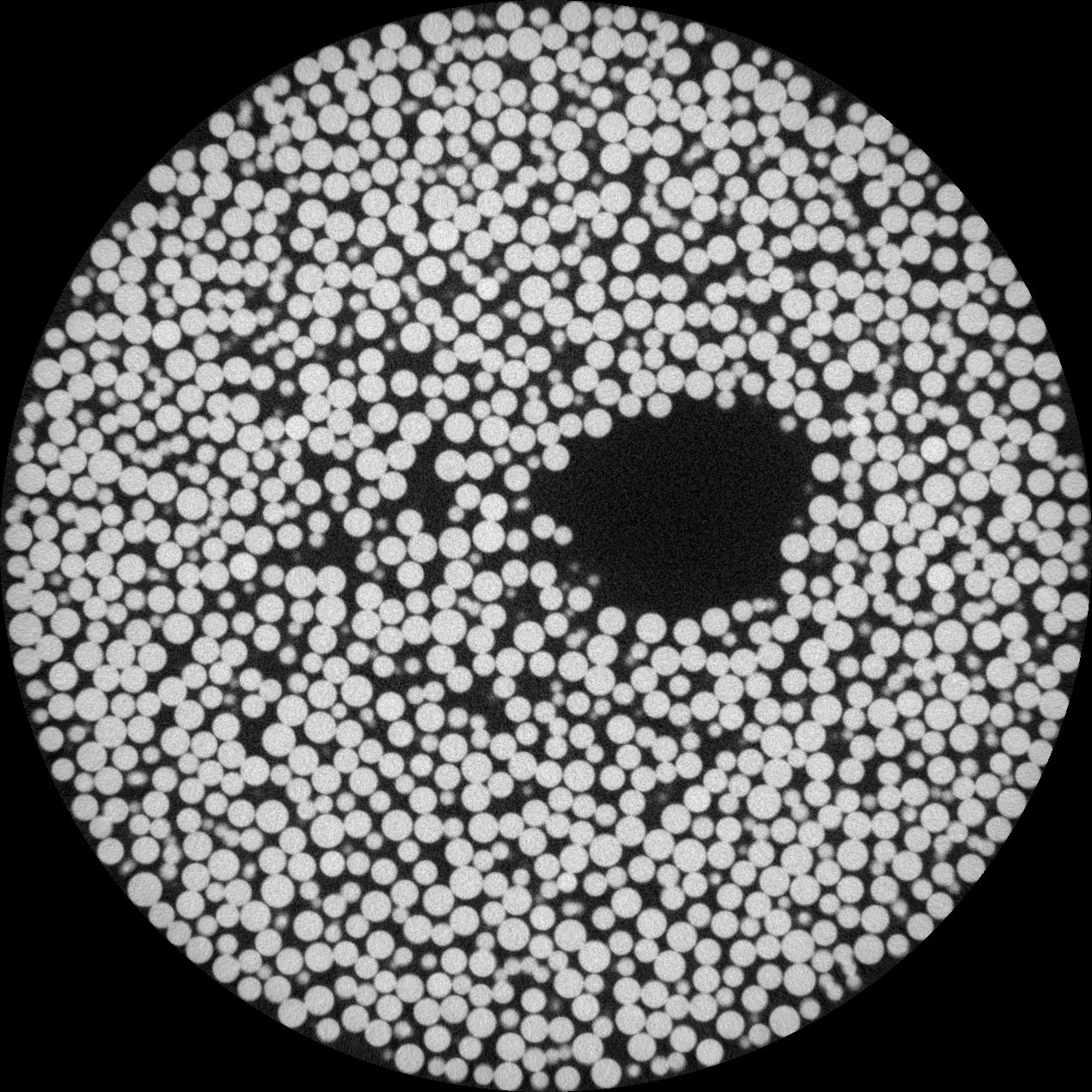} 
}
\qquad \qquad
\subfloat[$\mu$CT of vug space in core VC06 (x,y,z-resolution = 25 $\mu$m).]{%
\includegraphics[width=5cm]{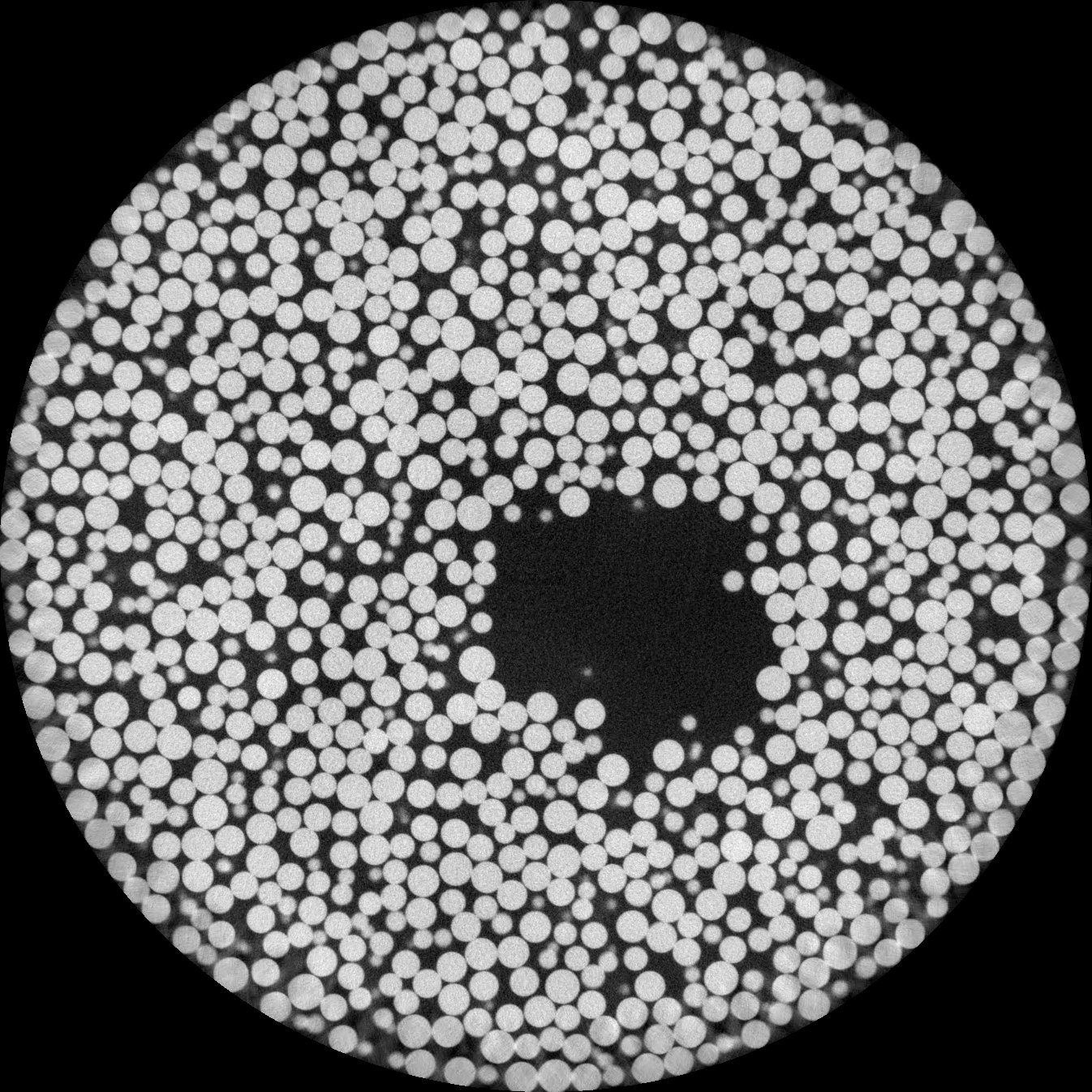}
}
\caption{The core fabrication process results in clear and well defined vug(s): no salt or cheese cloth residue are found.}
\label{fig:vug}
\end{figure}

Here we fabricate multiple glass bead cores with different vug heterogeneities (Figure \ref{fig:vugconfig}) using the temperature profile with a peak temperature of 725 $\degree$C and exposure time of 15 minutes (Figure \ref{fig:tempandcore}). The matrix porosity values is measured as $\sim$41\% (Figure \ref{fig:NMRhomo}) and can be altered by varying the peak temperature and exposure time.  Micro-CT scan across the middle of the vug for two different cores (Figure \ref{fig:vug}) show well-defined vug boundaries and no salt/cheesecloth residue.

\subsection{Core-fluid petrophysical properties} \label{corefluidprop}

The petrophysical properties for all the fabricated cores are determined experimentally. The methods are outlined below and the petrophysical values for all the cores are presented in Table \ref{tab:porocomp}.

\begin{table} \centering
\caption{\label{tab:porocomp}
Experimentally determined physical and petrophysical properties for each core.}
\begin{tabular}{c c c c c}  \centering 
&&&&\\ 
Core No & $\bar{\phi}_{CT}$ (\%) & $k_{liquid}$ (Darcy) & Vug vol. (frac.) & $\bar{r}_{pore}$ ($\mu m$) \\ 
\hline
 VC01 &  41.0 & 978 & 0.024 & 384 \\
 VC03 & 50.2 & 12417 & 0.109 & 1477 \\
 VC04 & 44.9 & 420 & 0.039 & 203 \\
 VC06 & 43.2 & 692 & 0.021 & 291 \\
 VC08 & 43.9 & 404 & 0.014 & 203 \\
 VC09 & 52.9 & 430 & 0.049 & 169 \\ 
 VC10 & 53.8 & 456 & 0.032 & 171 \\ \hline 
 HC01 & 41.6 & 100 & NO VUG & 97.8 \ddag \\
\hline
\end{tabular}
\\
{\footnotesize \ddag~{\itshape Figure \ref{fig:NMRhomo}}}
\end{table}

\paragraph{Porosity using computerized tomography (CT) imaging}

Core porosity is determined using an in-house modified medical computed tomography scanner from Universal Systems (HD-350E). The core is 100\% saturated with DI water, placed in the CT scanner, and scanned at a resolution of 250 $\mu$m. The scan parameters used are: 3 seconds scan time, 3 mm scan thickness, 3 mm scan point distance, 100 keV voltage, and 100 mA current. Since the core volume consists of only glass beads and DI water, the porosity can be determined by taking a weighted average (Eq. \ref{eqn:phi}) of the individual CT values for these two substances. The CT number for DI water and glass beads for this machine have been measured to be 25 HU and 2500 HU respectively. 

\begin{equation}
\phi = \frac{CT_{glass} - CT_{total}}{CT_{glass} - CT_{water}}
\label{eqn:phi}
\end{equation}

\begin{figure} \centering
\subfloat[Original CT image (left), histogram (center), and segmented image (right) using the Otsu method. \label{fig:seg}]{%
\includegraphics[height=4cm]{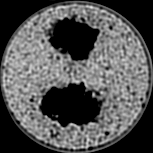}
\quad
\includegraphics[height=4cm]{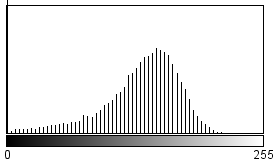}
\quad
\includegraphics[height=4cm]{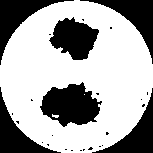}
}

\subfloat[3D visualization of the disconnected parallel vug system. \label{fig:vugvis}]{%
\includegraphics[width=7cm]{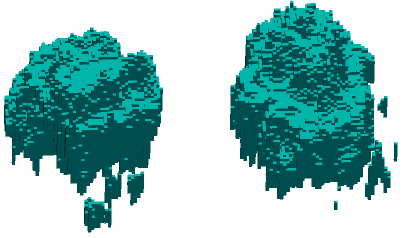}
}
\caption{Medical CT image, segmentation process, and 3D visualization of core VC09.}
\label{fig:VC09}
\end{figure}

The CT output is segmented (Figure \ref{fig:seg}) using the Otsu method \citep{otsu_threshold_1979} and the vug is isolated by using a connected region algorithm in ImageJ-Fiji \citep{schindelin_fiji:_2012,schneider_nih_2012}.The two parallel vug surface are visualized in Figure \ref{fig:vugvis}.

\paragraph{Single phase permeability using constant flow injection}
Single-phase liquid permeability of the cores is determined by pumping DI water through the core at variable flow rate (40, 50, and 60 ml/min) and measuring the pressure drop along the core. The pressure drop is used to calculate the liquid permeability using single-phase Darcy{\textquotesingle}s law.

\paragraph{Average pore throat radius}
The pore distribution in the homogeneous core (Figure \ref{fig:NMRhomo}) is determined by using an in-house Oxford Instruments GeoSpec2 NMR scanner with the average pore size calculated as 97.8 $\mu$m. 

This scanner can measure a maximum pore size of $\sim$350 $\mu$m before which the bulk relaxivity of the water dominates. The vugs introduced in the core are generally larger than this, therefore we use a theoretical Pittman R$_{50}$ equation \citep{pittman_relationship_1992} (Eq. \ref{eqn:pittman}) to determine the average pore radius for the vuggy cores. 

\begin{equation}
log_{10} ~(R_{50})  =  0.778  +  0.626~ log_{10}~ (k)  -  1.205~ log_{10} ~(\phi)
\label{eqn:pittman}
\end{equation}

\paragraph{Contact angle} \label{contactangle}
Contact angle measurements are made using a static sessile drop contact angle using a goniometer. The air/water contact angle on a sintered glass bead sheet, fabricated in presence of cheese cloth, was found to vary between 48\degree -- 50\degree.

\paragraph{Surface tension}
The surface tension of the air/water system is experimentally determined by conducting capillary rise in fixed diameter tubing. Two plastic tubes, with inner diameters 2.367 mm and 1.605 mm, are placed vertically in DI water and the elevation of air/water interface from the top of the fluid level is measured via a 2D tomogram. The contact angle of the setup is determined from the tomogram and the surface tension (Eq. \ref{eqn:surfacetension}) for the air-water system is calculated as 51.2 mN/m.

\begin{equation}
    \sigma = \frac{\rho g \cdot h  R}{2  \cos{\theta}}
    \label{eqn:surfacetension}
\end{equation}

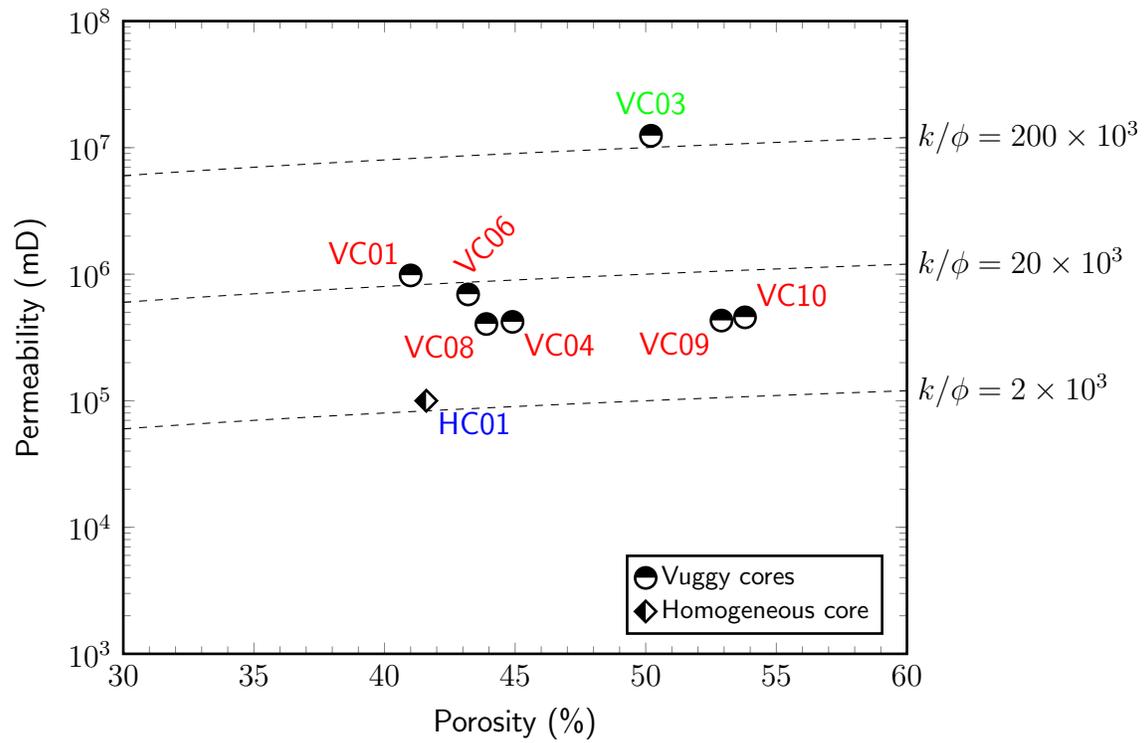
\begin{figure} \centering
\pgfplotsset{width=12cm,height=10cm,{every axis/.append style={line width=1pt}}}
\begin{tikzpicture}[font=\sffamily]
\begin{semilogyaxis}[xlabel=Porosity (\%),ylabel=Permeability (mD), 
	xmin=30, xmax=60, ymin=1e3, ymax=1e8,
	legend cell align=left, 
   	legend pos=south east,
	xtick distance=5, minor x tick num=4,clip=false]
	
\addplot [mark=halfcircle*,mark options={scale=2},only marks] table [y expr={\thisrow{K_Darcy}*1e3}, x=phi]  {Kphi_vug1.txt}; \addlegendentry{\footnotesize Vuggy cores}
\addplot [mark=halfsquare left*,mark options={draw=black,fill=black,scale=2}, only marks] table [x=phi, y expr={\thisrow{K_Darcy}*1e3}]  {Kphi_core16.txt}; \addlegendentry{\footnotesize Homogeneous core}

\node [red,above left] at (axis cs: 41,978e3)  {VC01};
\node [green,above] at (axis cs: 50.2,15000e3) {VC03};
\node [red,below right] at (axis cs: 44.9,420e3)  {VC04};
\node [red,above right,rotate=45] at (axis cs: 43.2,692e3) {VC06};
\node [red,below left] at (axis cs: 43.9,404e3) {VC08};
\node [red,below left] at (axis cs: 52.9,430e3)  {VC09};
\node [red,above right] at (axis cs: 53.8,456e3) {VC10};
\node [blue,below right] at (axis cs: 41.6,100e3) {HC01};

\addplot+[domain=30:60,black,dashed,no markers,thin] {2e3*x} node [pos=1,right] {$k/\phi = 2\times10^3$}; 

\addplot+[domain=30:60,black,dashed,no markers,thin] {20e3*x} node [pos=1,right] {$k/\phi = 20\times10^3$}; 
 
\addplot+[domain=30:60,black,dashed,no markers,thin] {200e3*x} node [pos=1,right] {$k/\phi = 200\times10^3$}; 

\end{semilogyaxis} 
\end{tikzpicture} 
\caption{Crossplot of experimentally determined permeability and total porosity for each core. All the cores have the same matrix porosity and pore size distribution. The k/$\phi$ contours are superimposed and flow units are identified by the text color (blue --  FU-1, red --  FU-2, and green --  FU-3).}
\label{fig:xplot}
\end{figure}

\paragraph{Porosity-permeability crossplot}
Porosity-permeability crossplot for the vuggy and homogeneous cores is plotted in Figure \ref{fig:xplot}, along with the k/$\phi$ ratio contours. The contours trace points with similar flow quality \citep{chopra_development_1987} and show three distinct flow units (FU) for the samples. FU-1 (colored blue in Figure \ref{fig:xplot}) has the lowest flow quality with a k/$\phi$ value of $2\times10^3$, FU-2 (red) has a k/$\phi$ value of $1.2\times10^4$, and FU-3 (green) has the best flow quality with a k/$\phi$ value of $2.5\times10^5$.

The presence of vug affects the flow unit classification of the core as observed via the departure of the points from the homogeneous core. The impact is more pronounced at the higher vug volume ratio: VC03 with a vug volume ratio $>$ 0.1 occupies FU-3 with a k/$\phi$ of 2.47$\times$10$^5$. Overall, we find that the spatial distribution and location of the vug does not have any impact on the flow unit classification. 



\section{Capillary rise } 

\subsection{Experiment methodology}\label{experiments}

Capillary rise over time is measured for all the carbonate proxy cores (Figure \ref{fig:vugconfig}): the core is placed in a Petri dish filled with DI water to a height of 10 mm and the water front movement is recorded by continuous optical imaging (Huawei Nexus 6P camera -- 12.3 MP Sony Exmor IMX377 with 4K video capture) at a frequency of 120 frames/second. The output video is converted to an image stack and binarized to determine the position of the air-water interface with time. 

\begin{figure} \centering
\includegraphics[width=0.9\textwidth]{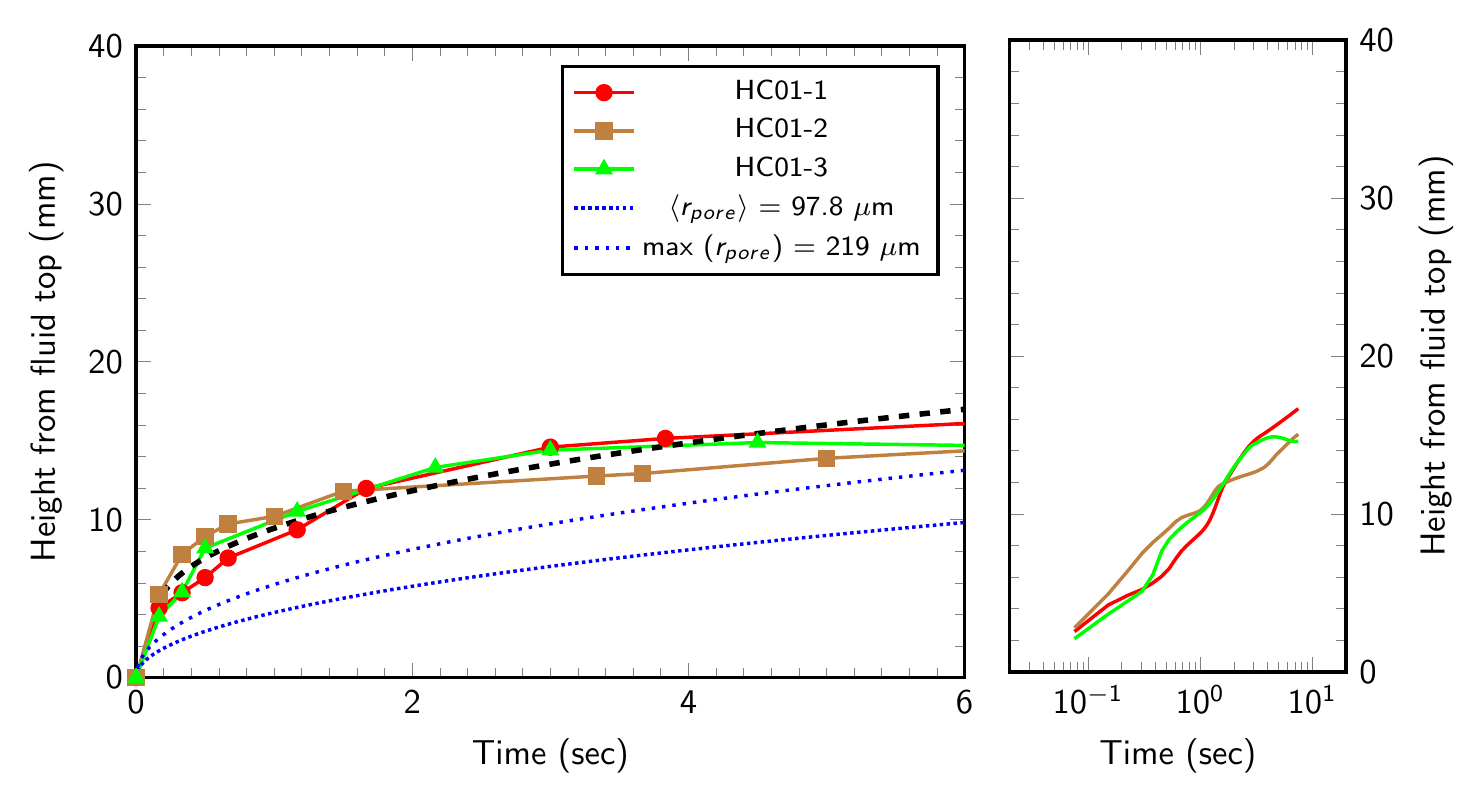}
\caption{Capillary rise (left) in three replicas of the homogeneous core (HC01) show repeatability of capillary rise. The experimental data is plotted along with the Fries-Dreyer solution to the Washburn equation for the average and maximum pore radius. The black line is a translation of the Washburn solution to the end of the initial sharp rise. Same results plotted on the log-log axis show an average gradient of 0.5.}
\label{fig:capriseHOMO}
\end{figure}

Three iterations of HC01 are produced and the capillary rise experiment is performed in each of them. A similar capillary rise profile is observed in all of them (Figure \ref{fig:capriseHOMO}): a sharp initial increase which lasts for the first 0.15 seconds followed by a logarithmic decline. The capillary rise is plotted along with the analytical solution for the Washburn equation in a grain pack \citep{fries_analytic_2008} for the experimentally determined maximum and average pore radii (Figure \ref{fig:NMRhomo}). The solution, which assumes a continuous and uniform porous medium bounded on the sides and is not true in this case, and is not able to capture the initial sharp fluid rise and ultimately under-predicts the total capillary rise. 

Compensating for the initial sharp fluid rise by transposing the solution to the end of that segment (black dashed line in Figure \ref{fig:capriseHOMO} left)  captures the curve nicely and results in a better match with the experimental observations. The capillary rise follows the Lucas--Washburn $\sqrt{t}$ law \citep{washburn_dynamics_1921} (Figure \ref{fig:capriseHOMO} right), which is consistent with previous studies in spontaneous imbibition in homogeneous porous media \citep{gruener_capillary_2009,gruener_anomalous_2012}.

\begin{figure} \centering
\includegraphics[width=0.9\textwidth]{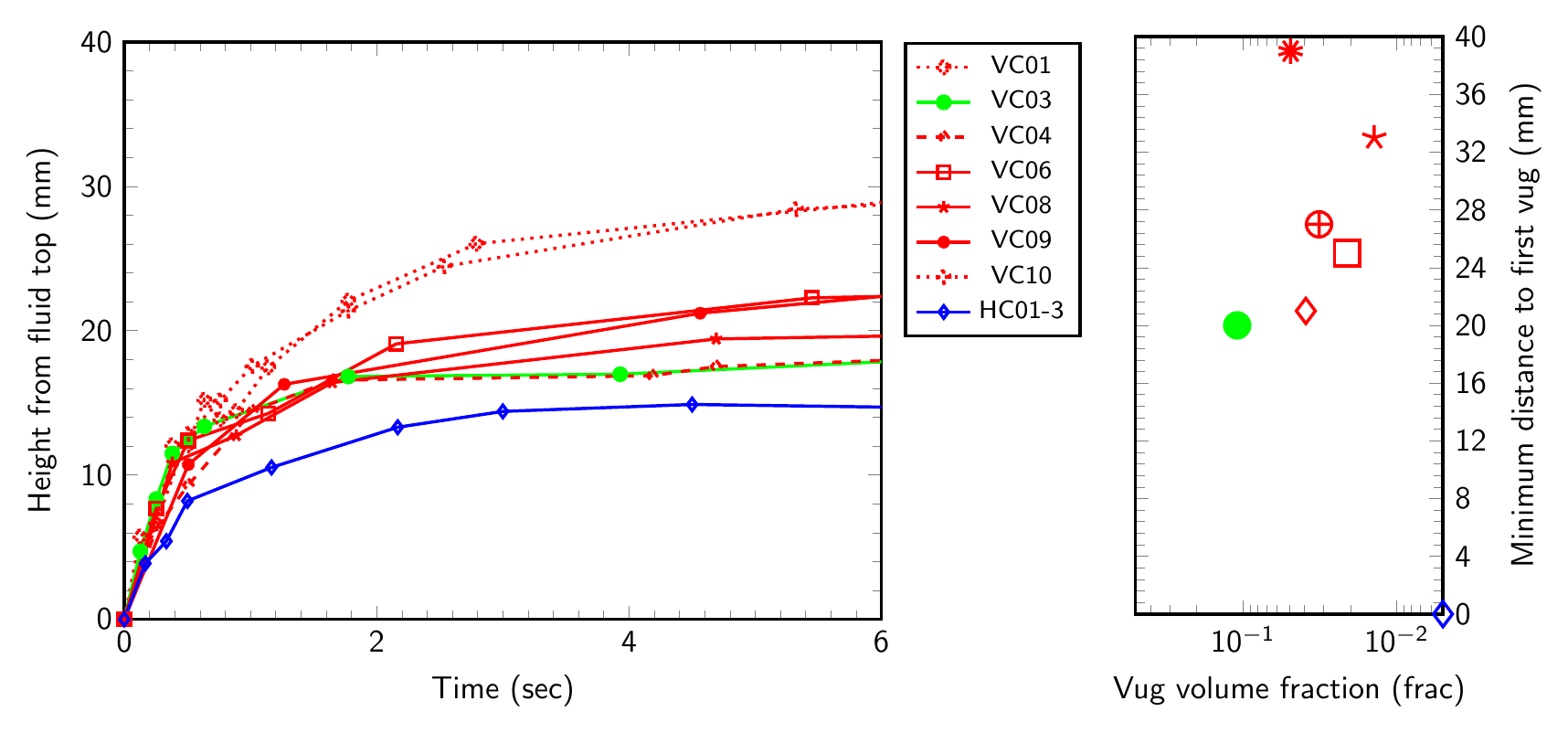}
\caption{Capillary rise (left) in all the vuggy cores along with one iteration of the homogeneous core are plotted against time. The plots are color-coded based on the flow unit they belong to (Figure \ref{fig:xplot}). The minimum distance to the bottom of the lowest vug is plotted on the right. Minimum distance to the vug for VC01 is outside the range for the plot, at 45 mm. The stabilized capillary rise does  not depend on the height of the bottom-most vug, with the fluid never touching the vug in some instances.}
\label{fig:capriseALL}
\end{figure}

Capillary rise in all the vuggy cores is plotted in Figure \ref{fig:capriseALL} along with one iteration of HC01. The line color denotes the flow unit the core belong to and the marker shape denotes the vug configuration. A large spread in the capillary rise for the vuggy cases is observed; all the vuggy cores result in a higher capillary rise than the homogeneous cores. The capillary rise can also be divided into two segments similar to the homogeneous core: a sharp initial fluid rise, followed by a logarithmic decline. The sharp initial fluid rise has the same gradient as HC01 and it lasts longer ($\sim$~0.45 seconds).

Figure \ref{fig:capriseALL}(right) shows the difference in elevation of the bottom of the first vug from the static water level for all the cores. The experiments show no discernable trend in the the spatial location of the vug (distance from static water level to the start of first vug) and the capillary height at equilibrium. The only observable fact is that the presence of vug results in a higher capillary rise in the porous medium; for some cases even before the fluid reaches the vug.

\subsection{Computational fluid dynamics simulation}
\begin{figure} \centering
\subfloat[Two-tube model: two tubes of equal diameter (1 mm) connected with a 0.1 mm diameter tubing at a height 3 mm above the static fluid level. \label{fig:twotube}]{%
\includegraphics[width=0.4\textwidth]{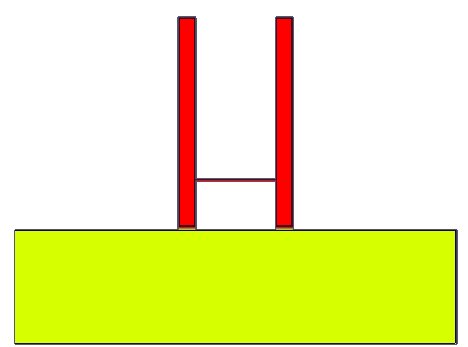}
}
\qquad
\subfloat[Tube-vug model: two tubes of equal diameter (1 mm) connected with a 0.1 mm diameter tubing at a height 3 mm above the static fluid level. The right tube funnels from 1 mm diameter at 7 mm height to 6 mm diameter at 13 mm above fluid level.\label{fig:tubevug}]{%
\includegraphics[width=0.4\textwidth]{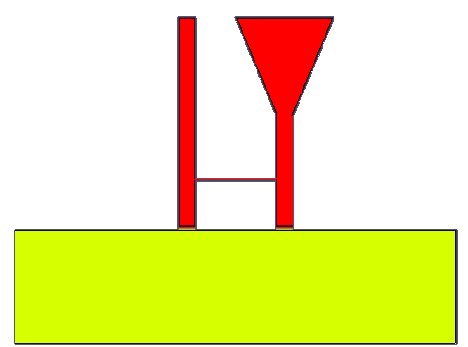}
}
\caption{The color in the two models (two-tube and tube-vug model) denote the ambient fluid saturation ($\alpha_{water}$) at initial time: fluid reservoir in yellow ($\alpha_{water}$ = 1) and the dry tubes ($\alpha_{water}$ = 0) being red. The fluid reservoir is significantly larger than the combined volume of the tubes, with the tubes dipping 2 mm below the fluid top.  }
\label{fig:CFDsetup}
\end{figure}

Capillary rise was modeled by numerically solving the Navier-Stokes equations for two incompressible isothermal immiscible Newtonian fluids using the InterFOAM solver in OpenFOAM\textsuperscript{\textregistered} \citep{weller_tensorial_1998}. We consider two scenarios: (i) two connected tubes of equal diameter (Figure \ref{fig:twotube}); and (ii) two connected tubes of equal diameter till 7 mm above fluid level, and then increasing linearly six-fold at 13 mm above fluid level (Figure \ref{fig:tubevug}). The height of the two tubes and the height of fluid above the tube entrance, for both the scenarios, is the same. All tubes are open to the atmosphere and have a constant (atmospheric) pressure boundary. The fluid reservoir is open to the atmosphere, with the top boundary at atmospheric pressure. A high resolution mesh is required to simulate the capillary rise phenomenon. The input parameters are shown in Table \ref{table:OFtable}. The surface tension and contact angle at the air-fluid interface is selected based on the experimental measurement (section \ref{corefluidprop}). 

\begin{table}
\centering \caption{OpenFOAM\textsuperscript{\textregistered} input parameters.}

\begin{tabular}{| c | c|}
\hline
 \bfseries{Input parameter} & \bfseries{Value} \\ \hline
Contact angle & 49$\degree$ \\ \hline
Surface tension & 0.0512 N/m  \\ \hline
Air density & 1 kg/m$^3$ \\ \hline
Air viscosity & $1.48\times10^-5$ Pa.s \\ \hline
Water density & 1000 kg/m$^3$ \\ \hline
Water viscosity & $1\times10^-3$ Pa.s\\ \hline

\end{tabular} 
\label{table:OFtable}
\end{table}

\begin{figure} \centering
\pgfplotsset{width=0.45\textwidth,height=7cm,{every axis/.append style={line width=1pt}}}
\subfloat[Capillary rise in left tube for both models. \label{fig:CFDleft}]{%
\begin{tikzpicture}
\begin{axis}[xlabel=Time (sec), ylabel=Capillary rise from static fluid level (mm), 
	xmin=0, xmax=2, ymin=00, ymax=10,
	legend style={cells={align=left}}, 
   	legend pos= north west,xtick distance=0.5,
   	xminorticks=true, minor x tick num= 9,
   	yminorticks=true, minor y tick num=3,]

\addplot [red] table [x=Time, y=H_TwoTube]  {capillaryheight_leftsimulation.txt}; \addlegendentry{\footnotesize Two-tube model}
\addplot [blue,dashed] table [x index={0}, y index={2}]  {capillaryheight_leftsimulation.txt}; \addlegendentry{\footnotesize Tube-vug model}
\addplot [black,dotted,line width=1pt] {7};
\node at (axis cs:0.85,2) {\includegraphics[height=13mm]{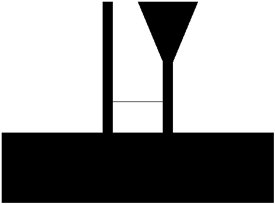}};
\addplot [patch,patch type=rectangle,mesh,blue] coordinates {(0.73,1) (0.85,1) (0.85,3.5) (0.73,3.5)};
\node at (axis cs:1.5,2) {\includegraphics[height=13mm]{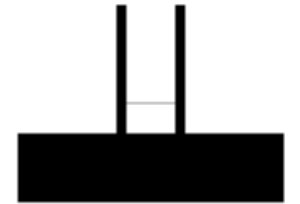}};
\addplot [patch,patch type=rectangle,mesh,red] coordinates {(1.38,1) (1.5,1) (1.5,3.5) (1.38,3.5)};

\end{axis} 
\end{tikzpicture} 
}
\subfloat[Capillary rise in right tube for both models. \label{fig:CFDright}]{%
\begin{tikzpicture}
\begin{axis}[xlabel=Time (sec), ylabel=Capillary rise from static fluid level (mm), 
	xmin=0, xmax=2, ymin=00, ymax=10,
	legend style={cells={align=left}}, 
   	legend pos= north west,xtick distance=0.5,
   	xminorticks=true, minor x tick num= 9,
   	yminorticks=true, minor y tick num=3,]

\addplot [red] table [x=Time, y=H_TwoTube]  {capillaryheight_rightsimulation.txt}; \addlegendentry{\footnotesize Two-tube model}
\addplot [blue,dashed] table [x index={0}, y index={2}]  {capillaryheight_rightsimulation.txt}; \addlegendentry{\footnotesize Tube-vug model}
\addplot [black,dotted,line width=1pt] {7};
\node at (axis cs:0.85,2) {\includegraphics[height=13mm]{tubevug_base.jpg}};
\addplot [patch,patch type=rectangle,mesh,blue] coordinates {(0.83,1) (1,1) (1,3.5) (0.83,3.5)};
\node at (axis cs:1.5,2) {\includegraphics[height=13mm]{twotube_base.jpg}};
\addplot [patch,patch type=rectangle,mesh,red] coordinates {(1.51,1) (1.64,1) (1.64,3.5) (1.51,3.5)};

\end{axis} 
\end{tikzpicture} 
}
\caption{Capillary rise in the left and right tubes for the two models. The dotted black line represents the height at which the tube diameter changes for the tube-vug model (Figure \ref{fig:tubevug}). The tube-vug model shows the air-fluid interface at a greater height in the left tube.}
\label{fig:capriseSIM}
\end{figure}

Capillary rise is simulated; the fluid level rises in both tubes and is proportional to the square root of time (Figure \ref{fig:capriseSIM}). Fluid preferentially flows in the connected tube (3 mm above static fluid level) until it is filled, which can be observed by the short-duration departure from the proportional relationship at $\sim$3 mm fluid height. Subsequently, the fluid level continues to increase proportional to the square root of time in both tubes for the two-tube model. 

The fluid level in the right tube does not increase after 7 mm (above static fluid level) for the tube-vug model, where the tube diameter starts increasing. On the other hand the fluid level in the constant diameter left tube keeps on increasing with time, keeping the same shape as the two-tube model. Deviation between the two models starts at 1.60 seconds, with the fluid level in the tube-vug model rising quicker (Figure \ref{fig:capriseSIM}).


\subsection{Micro-tomographic imaging}

\begin{figure} \centering
\subfloat[Air-water interface at equilibrium for a continuous glass bead proxy core at a resolution of 24 $\mu$m. \label{fig:expriseCONT}]{%
\includegraphics[height=0.33\textwidth]{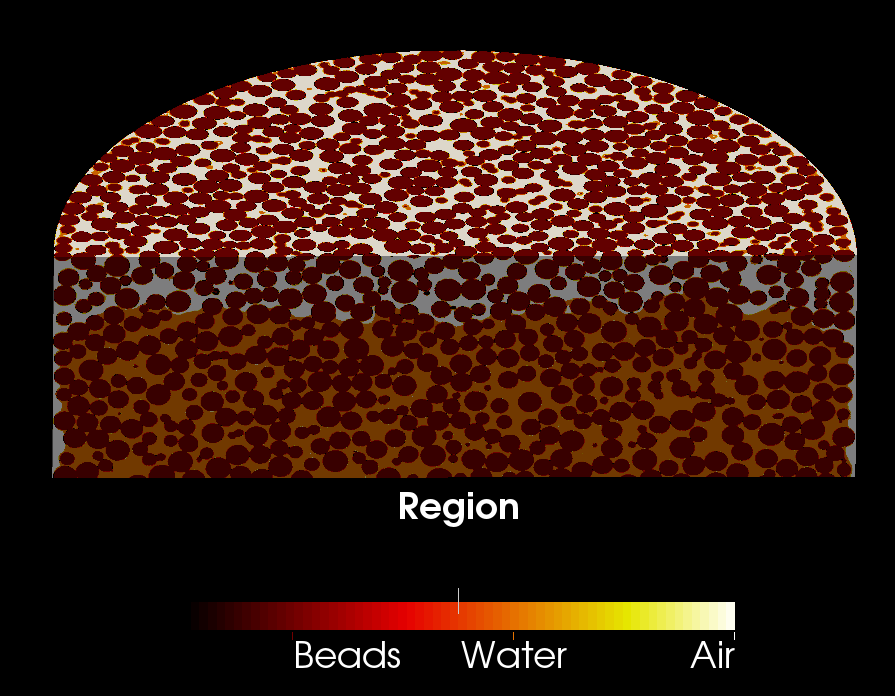}
}
\quad
\subfloat[Air-water interface at equilibrium for a vuggy glass bead proxy core at a resolution of 25 $\mu$m. \label{fig:expriseVUG}]{%
\includegraphics[height=0.33\textwidth]{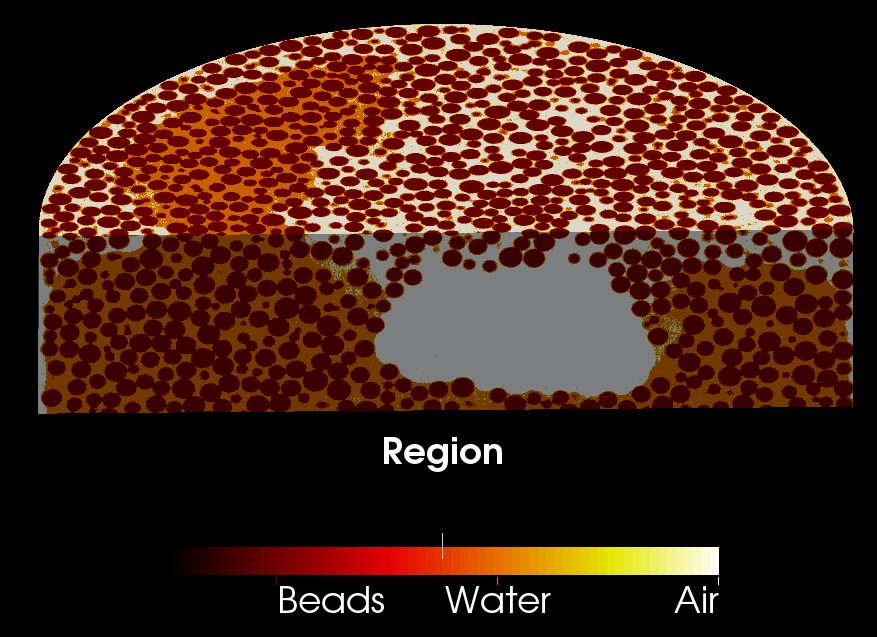}
}
\caption{False-colored segmented 3D reconstruction of air-water interface in a vuggy and a homogeneous proxy core. Air is colored white, glass beads are colored red and water is colored orange.}
\label{fig:uCTriseExp}
\end{figure}

Capillary rise on a homogeneous (HC01) and a vuggy core (VC06) is performed inside the in-house x-ray micro-tomography scanner. Once the fluid front stabilizes, the cores are scanned at a resolution of 25 $\mu$m to get the spatial fluid distribution inside the core (Figure \ref{fig:uCTriseExp}). The scans are segmented to identify the three components: air (white), water (orange), and solid glass (red), and a 3D volume is reconstructed. 

The 3D reconstruction of the air-water interface for HC01 is shown in Figure \ref{fig:expriseCONT}. The air-water interface, at the boundary of the white and orange colors, is fairly uniform across the cross-section with minor variations (less than one bead diameter (1.0 mm)). These variations can be attributed to the local variation in bead packing, and in effect, the pore structure. The air-water interface near the outer surface of the cores has a curvature to it and results in a lower water height at the core boundary. 

A similar 3D reconstruction of the air-water interface for the vuggy core (VC06) is shown in Figure \ref{fig:expriseVUG}, where the air-filled vug is colored white. The air-water interface for this core configuration is not uniform across the cross-section with large variations (larger than one bead diameter (1.0 mm)) observed across it. The water level is generally lower near the vug-matrix boundary and the core-atmosphere boundary; the water level increases when it is further away from these boundaries resulting in a higher water level in the matrix with the thicker porous medium. The air-water interface near the outer surface of the core has a curved surface, similar to one observed in HC01, with the water level lower at the core external boundary.

\section{Discussion}
Combining the above observations, we can clearly see that the presence of disconnected vugs result in a higher capillary rise. The vug acts as a capillary barrier to the imbibing fluid, diverting the fluid to a connected alternative path, which ultimately results in a higher capillary rise.

\begin{figure} \centering
\pgfplotsset{width=10cm,height=6cm,{every axis/.append style={line width=1pt}}}
\begin{tikzpicture}
\begin{axis}[xlabel=Time (sec), ylabel={Distance to interface from \\ fluid top (mm)}, ylabel style={align=center}, xmin=0, xmax=0.6, ymin=00, ymax=20,legend pos=outer north east, xtick distance=0.2, xminorticks=true, minor x tick num = 3, yminorticks=true, minor y tick num=4]

\addplot [red,dotted] table [x index={0}, y index={1}] {VC01.txt}; \addlegendentry{\footnotesize VC01}
\addplot [green] table [x index={0}, y index={1}] {VC03.txt}; \addlegendentry{\footnotesize VC03}
\addplot [red,dashed] table [x index={0}, y index={1}] {VC04.txt}; \addlegendentry{\footnotesize VC04}
\addplot [red,loosely dashed] table [x index={0}, y index={1}] {VC06.txt}; \addlegendentry{\footnotesize VC06}
\addplot [red,dashdotted] table [x index={0}, y index={1}] {VC08.txt}; \addlegendentry{\footnotesize VC08}
\addplot [red,densely dashed] table [x index={0}, y index={1}] {VC09.txt}; \addlegendentry{\footnotesize VC09}
\addplot [red] table [x index={0}, y index={1}] {VC10.txt}; \addlegendentry{\footnotesize VC10}
\addplot [blue] table [x index={0}, y index={1}]  {Cont2B.txt}; \addlegendentry{\footnotesize Homogeneous}

\end{axis} 
\end{tikzpicture} 

\caption{Zoomed-in version of Figure \ref{fig:capriseALL}. The duration of initial capillary rise is different for each flow unit.}
\label{fig:experimentalresultszoom}
\end{figure}
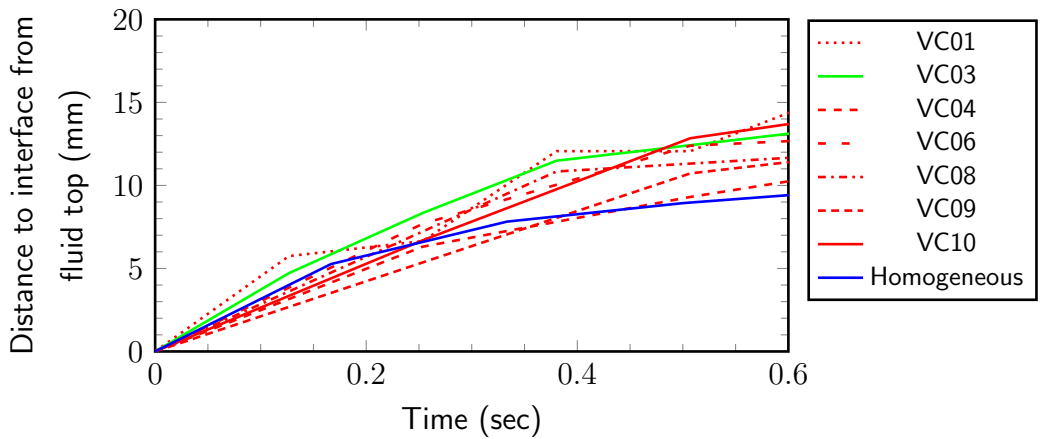

The capillary rise behavior for all the cores can be split into two zones as discussed in Section {\ref{experiments}}: (i) an initial quick rise; and (ii) an exponential decay. All the cores agree on the rate of initial capillary rise (Figure \ref{fig:experimentalresultszoom}), though the duration for the homogeneous core ($\sim$~0.15 seconds) is considerably shorter than the average duration for the vuggy cores ($\sim$~0.45 seconds). The similar rate of increase can be explained by the large permeabilities for all the cores which results in a short degassing time (time for air pressure to reduce from atmospheric pressure to zero in the core), on the order of $1\times10^{-3}$ seconds, which is considerably smaller than the duration of initial rise. The initial rise duration for the vuggy cores can be grouped according to their flow unit classification (Figure \ref{fig:xplot}): FU-1 show the longest duration while FU-3 shows the shortest. Permeability difference between the vuggy cores (Table \ref{tab:porocomp}) explains the difference in initial rise duration; a better flow quality, i.e. FU-3, starts stabilizing quicker and would therefore have the shortest duration.  

Capillary rise at equilibrium is not dependent on the flow unit type (Figures \ref{fig:experimentalresultszoom}). Two generalization can be made: (i) the presence of vug(s) results in a higher capillary rise at equilibrium conditions; and (ii) the highest capillary rise (VC01 and VC10) is in the core with the largest distance to the top of the furthest vug from the bottom. A similar result is also obtained from a CFD simulation (Figure \ref{fig:capriseSIM}), where the fluid attains a higher capillary rise in the connected tube-vug model compared to the two-tube model. We deduce that the momentum of the fluid in the vuggy tube (right tube in Figure \ref{fig:tubevug}) is transferred to the constant diameter tube (left tube in Figure \ref{fig:twotube}), which pushes the fluid to a greater height and shows the departure from the two-tube model observed in Figure \ref{fig:CFDleft}. A higher equilibrium height is also observed in $\mu$CT scans (Figure \ref{fig:uCTriseExp}), which also shows the fluid by-passing the vug and preferentially rising on one side (in an uneven manner). 

\begin{figure} \centering
\includegraphics[width=3.5cm]{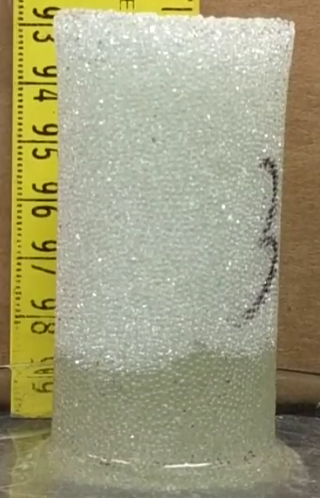}
\qquad
\includegraphics[width=6cm]{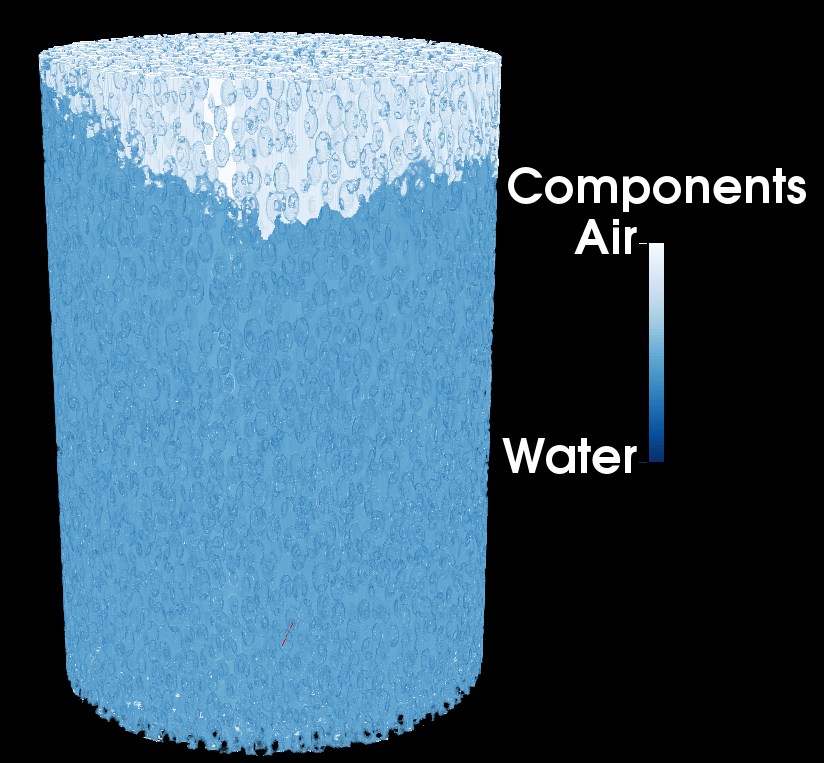}

\caption{The presence of a vug results in a non-uniform fluid front across the width of the core (left -- experiment and right -- microtomogram) which leads to error in determining the true capillary rise value.}
\label{fig:frontunevenwithporosity}
\end{figure}

A point to take into consideration is that the fluid front is not uniform across the cross-section for vuggy core (Figure \ref{fig:frontunevenwithporosity}), and for our experiments we have taken the maximum fluid height as the capillary rise value at each timestep. Furthermore, the optical imaging is performed from one side of the core; it is possible that the capillary fluid front might be present at a different height when observed from the backside (Figure \ref{fig:expriseVUG}). 

For most cases, vug(s) impact the capillary rise even before the fluid interacts with them (Figure \ref{fig:capriseALL} right). This suggests a feedback mechanism that transmits information of the high permeability ahead, even before the fluid front reaches that point. 

\begin{figure} \centering
\subfloat[Reduction in the matrix cross-sectional area results in a higher velocity of the fluid, with the jetting action pushing it forward.]{%
\includegraphics[width=5cm]{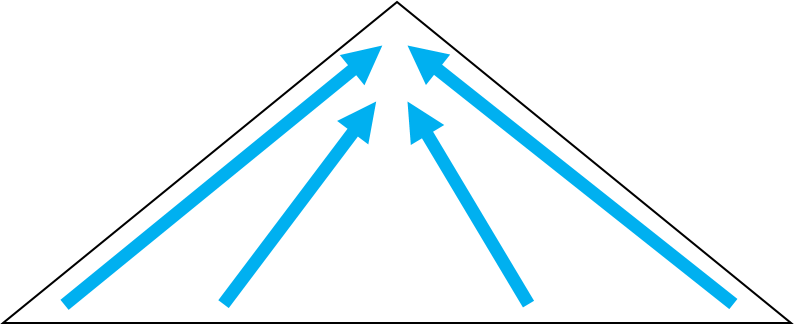}
}
\qquad
\subfloat[Once the vug is passes, the matrix cross-sectional area increases which reduces the velocity of the fluid and slows down the fluid movement.]{%
\includegraphics[width=5cm]{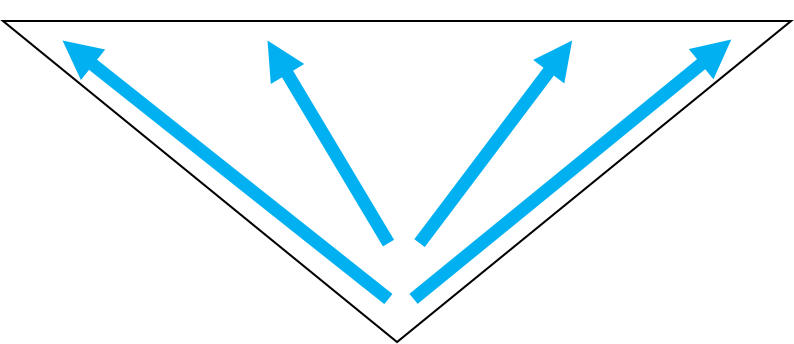}
}
\caption{Converging and diverging geometry of the matrix affect the fluid front movement.}
\label{fig:ConvDivFluid}
\end{figure}

\begin{figure} \centering
\subfloat[3D reconstruction of water (white) surrounding and rising around the vug (blue). It does not permeate inside the vug. \label{fig:vugencap1}]{%
\includegraphics[height=6cm]{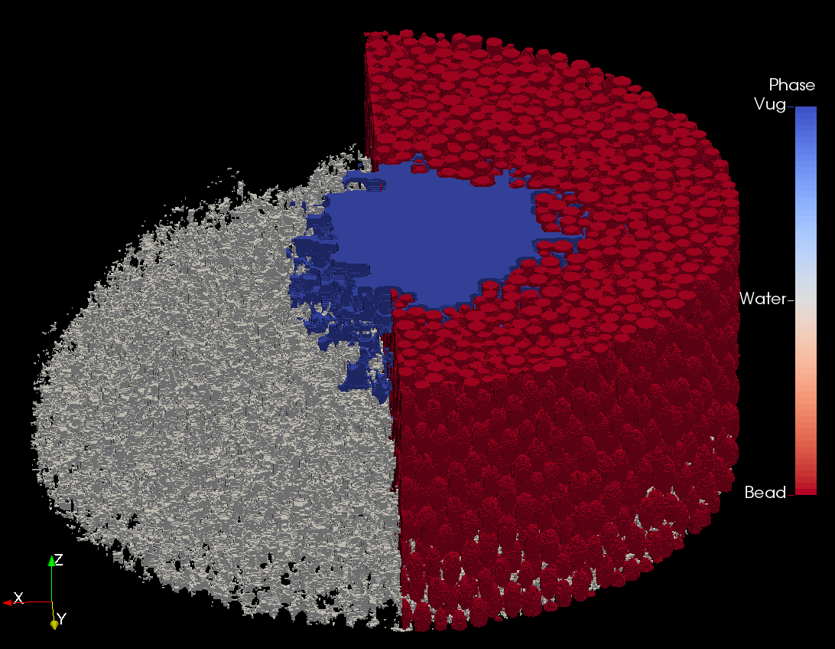}
}
\qquad
\subfloat[The vug is filled with air (white) and is completely encapsulated by the water (orange) phase. The air-water interface is not horizontal, with a higher water level at the edges. \label{fig:vugencap2}]{%
\includegraphics[height=6cm]{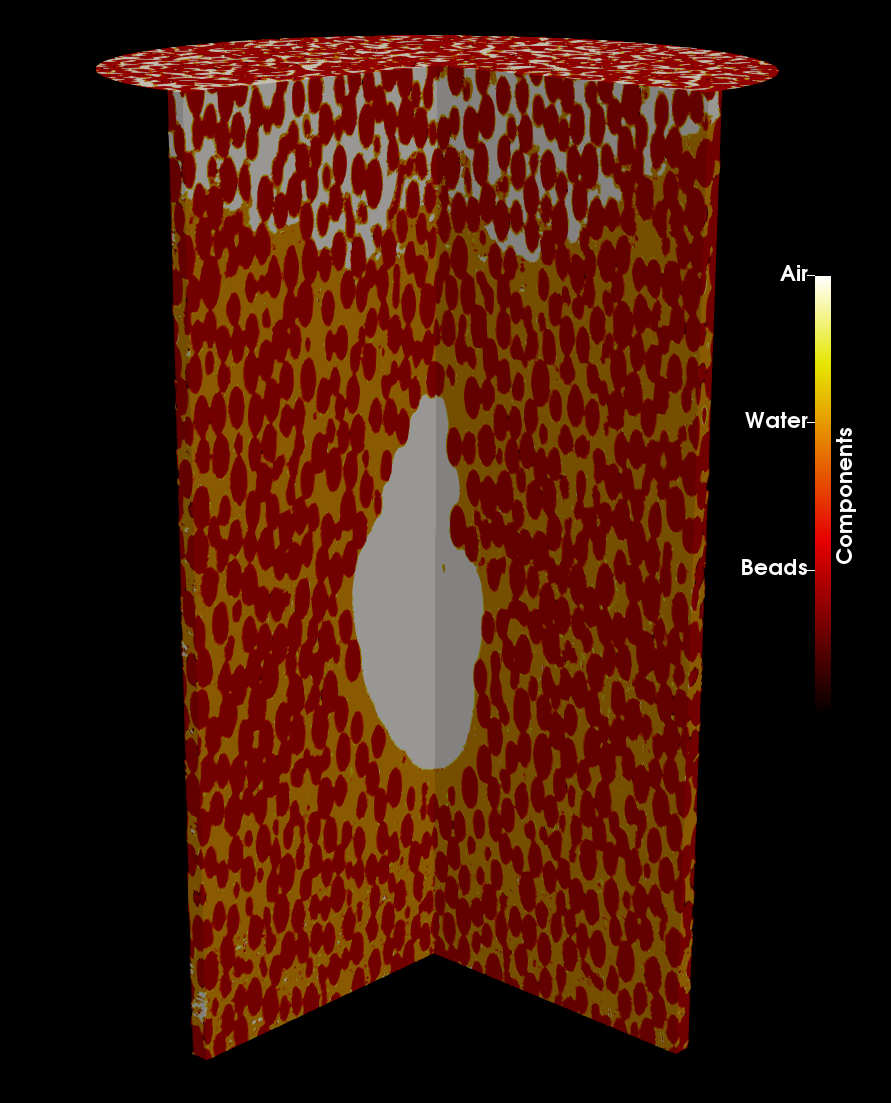}
}
\caption{The water phase requires a higher capillary pressure to enter the vug, which it does not possess, and therefore rises on the sides of the vug.}
\label{fig:VugEncap}
\end{figure}

Considering the change in matrix cross-sectional area within the presence of the vug when moving up along the length of the core, the fluid velocity will vary in the matrix (Figure \ref{fig:ConvDivFluid}). The converging geometry in the matrix, when moving towards the vug, results a higher fluid velocity which results in a jetting movement of the fluid in the upward direction. As soon as the center of the vug is passed, the geometry becomes divergent and the fluid slows down. If the fluid velocity is large enough when it passes the vug, the fluid can completely encapsulate the vug (Figure \ref{fig:VugEncap}).

\section{Summary}

In sum, our results suggest that for a carbonate rock in the presence of a disconnected vug, the fluid (contaminant) penetrates deeper in the porous media compared to a homogeneous porous media.The disconnected vugs are not filled and fluid by-passes them as the momentum of the fluid in the capillary leading to the vug is transferred to a capillary that is circumventing the vug, which pushes the fluid to a greater height.The disconnected vugs affect the spontaneous imbibition behavior in a permeable media even before the fluid interacts with them. 

\section*{Acknowledgments}
This project started as part of The University of Texas at Austin Hildebrand Department of Petroleum and Geosystems Engineering (UT PGE) Summer Undergraduate Research Internship (SURI) program and continued under support from Digital Rocks Petrophysics Industrial Affiliates Program (PI Prodanovic).

\section*{Nomenclature}

\begin{tabular}{c|l}
$\phi$ & Porosity (\%) \\
$\theta$ & Contact angle (\degree) \\
$\rho$ & Density (g/ml) \\
CT & Computed tomography \\
DI & De-ionized \\
fps & Frames per second \\
FU & Flow unit \\
HC & Homogeneous core \\
HU     &  Hounsfield units \\
k & Permeability (D) \\
keV     & kilo electron volts \\
NMR & Nuclear Magnetic Resonance \\
R$_{50}$ & 50$^{th}$ percentaile pore radius ($\mu$m) \\
VC & Vuggy core \\ 
\end{tabular}

{\scriptsize
\bibliographystyle{unsrtnat}
\bibliography{ref}
}

\appendix
\renewcommand\thefigure{A.\arabic{figure}}    
\setcounter{figure}{0}    
\renewcommand\thetable{A.\arabic{table}}    
\setcounter{table}{0}

\section{Additional experiments}
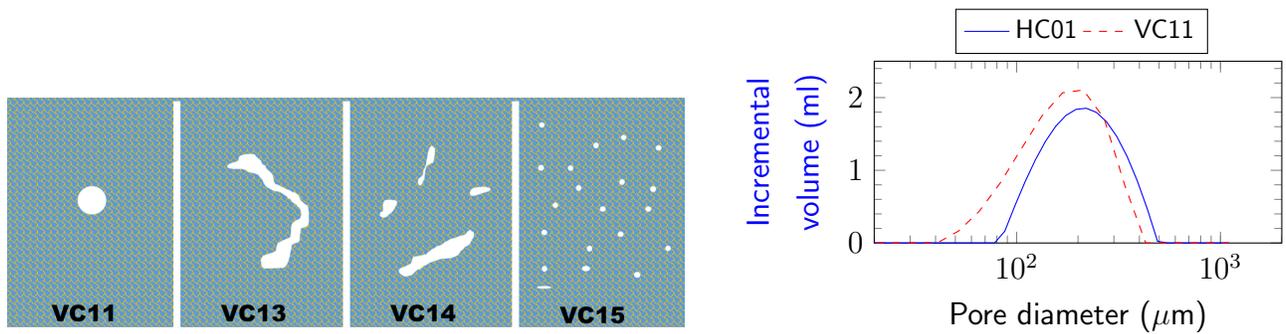
\begin{figure}
\centering
\includegraphics[width=9cm]{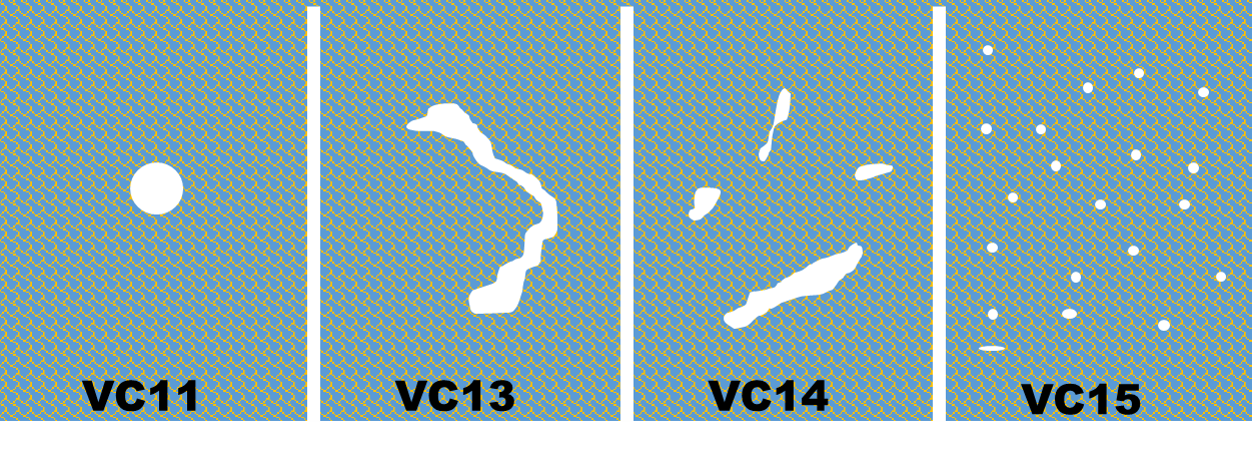}
\quad
\pgfplotsset{width=7cm,height=4cm}
\begin{tikzpicture}[font=\sffamily]
\begin{axis}[xlabel={Pore diameter ($\mu$m)} ,ylabel=\textcolor{blue}{Incremental volume (ml)},ylabel style={text width=2.5cm,align=center}, xmode=log, xmin=20, xmax=2000, ymin=0, ymax=2.5,minor y tick num=4,legend cell align=left, legend style={at={(axis cs:50,2.6)},anchor=south west},legend columns=-1,]
\addplot [blue] table [x index={0}, y index={1}]  {NMR.txt};  \addlegendentry{\footnotesize HC01}
\addplot [red,dashed] table [x index={0}, y index={1}]  {NMR_Core11.txt}; \addlegendentry{\footnotesize VC11}
\end{axis}  
\end{tikzpicture}
\caption{Vug configration (left) with the new temperature profile. NMR shows a lower average pore size in the matrix for this temperature profile.}
\label{fig:coreconfig2}
\end{figure}

The matrix in the vuggy core is changed by altering the temperature profile (Figure \ref{fig:tempandcore}): increasing the peak temperature to 775 $\degree$C and exposure time to 25 minutes. This results in a matrix porosity of 19.5\%. Four new cores with varying different vug configurations (Figure \ref{fig:coreconfig2}) are fabricated. Petrophysical properties are experimentally determined (Table \ref{tab:coreprop}). When plotted on the porosity-permeability crossplot, these cores align with FU-1 (Figure \ref{fig:xplot}) with k/$\phi$ = 2$\times$10$^3$, which is similar to the homogeneous core (HC01) for the previous temperature profile. 

\begin{table} \centering
\caption{\label{tab:coreprop} Experimentally determined physical and petrophysical properties for the four cores.}
\begin{tabular}{c c c c c}  \centering 
&&&&\\ 
Core No & $\bar{\phi}_{CT}$ (\%) & $k_{liquid}$ (Darcy) & Vug vol. (frac.) & $\bar{r}_{pore}$ ($\mu m$) \\ \hline
 VC11 & 31.0 & 90 & 0.047 & 127 \\
 VC13 & 24.1 & 63 & 0.044 & 132 \\
 VC14 & 19.2 & 43 & 0.011 & 129 \\
 VC15 & 23.5 & 107 & 0.016 & 188 \\ \hline
\end{tabular}
\end{table}

\begin{figure}[!] \centering
\includegraphics[width=0.9\textwidth]{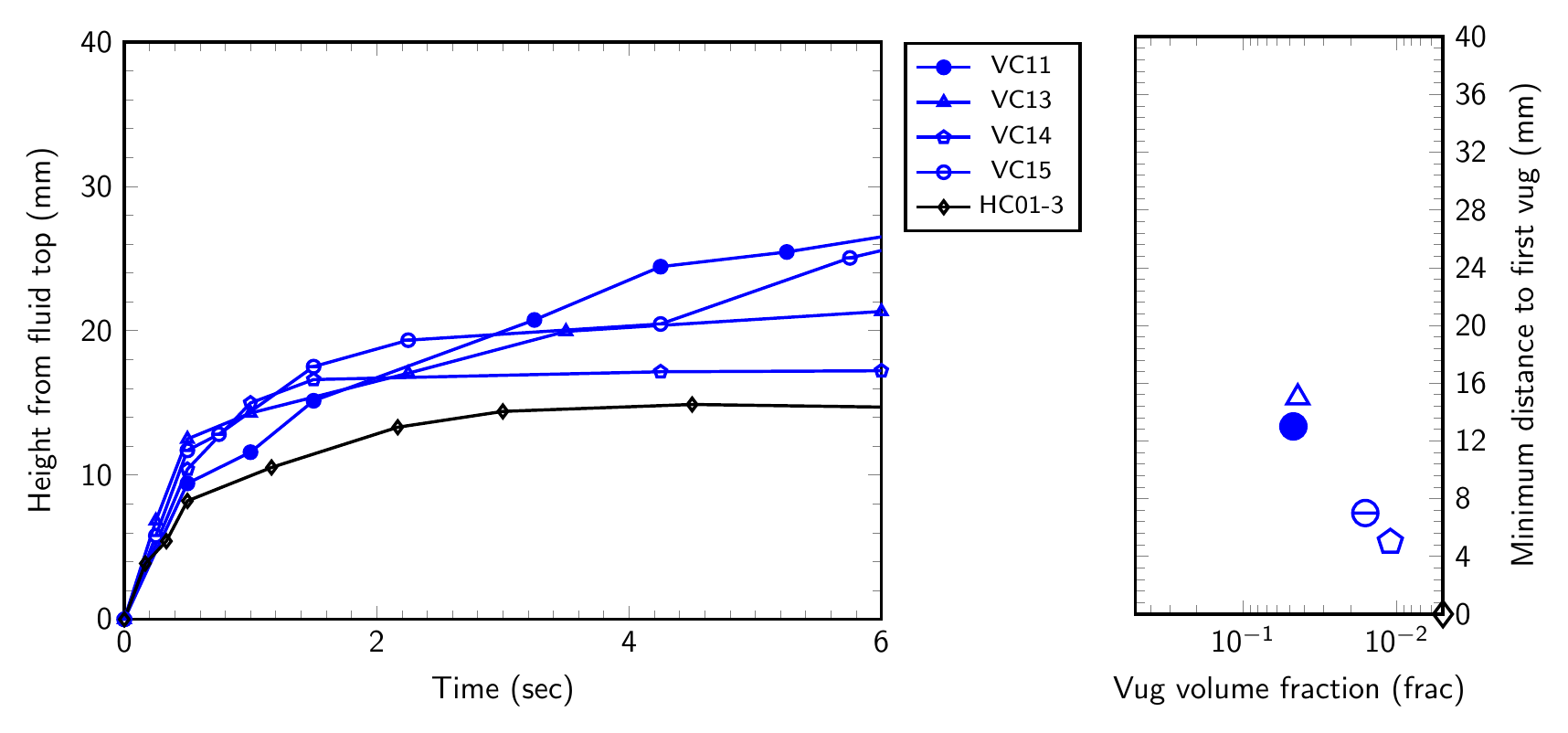}
\caption{Capillary rise (left) in all the vuggy cores (blue) along with one iteration of the homogeneous core (HC01 -- black) are plotted against time. The minimum distance to the bottom of the lowest vug is plotted on the right. All the core are part of FU-1. The stabilized capillary rise does  not depend on the height of the bottom-most vug, with the fluid never touching the vug in some instances.}
\label{fig:capriseNEW}
\end{figure}

Capillary rise is conducted in the same way (section \ref{experiments}) and the results are presented in Figure \ref{fig:capriseNEW} along with the capillary rise in one iteration of the homogeneous core (HC01) generated with the previous temperature profile. As observed for the previous cores, the capillary rise is in all the vuggy cores is higher than the rise in the homogeneous core. No discernable pattern is evident when comparing the capillary rise between the two temperature profiles (Figures \ref{fig:capriseALL} and \ref{fig:capriseNEW}).

\end{document}